\documentclass[iop]{emulateapj}
\usepackage{amsmath}

\newcommand\xone{x_1}
\newcommand\Farm{\mathcal{F}}

\newcommand\Oms{\Omega_{\rm arm}}
\newcommand\Omb{\Omega_{\rm bar}}
\newcommand\SFR{{\rm SFR}}
\newcommand\aSFR{\langle\SFR\rangle}
\newcommand\Rring{R_{\rm ring}}
\newcommand\Mcrit{\dot M_{\rm *, CP}}
\newcommand\Mring{M_{\rm ring}}
\newcommand\aMring{\langle\Mring\rangle}
\newcommand\RCRa{R_{\rm CR, arm}}
\newcommand\RCRb{R_{\rm CR, bar}}

\newcommand\Msun{\; {\rm M}_{\odot}}
\newcommand\erg{\; {\rm erg}}
\newcommand\kms{\; {\rm km}\;{\rm s}^{-1}}
\newcommand\pc{\;{\rm pc}}
\newcommand\kpc{\;{\rm kpc}}
\newcommand\freq{\;\kms\kpc^{-1}}

\newcommand\yr{\; {\rm yr}}
\newcommand\Myr{\;{\rm Myr}}
\newcommand\Gyr{\;{\rm Gyr}}
\newcommand\Aunit{\Msun\yr^{-1}}
\newcommand\Surf{\Msun\pc^{-2}}
\newcommand\DSN{\Delta \tau_{\rm SN}}
\newcommand\simgt{\lower.5ex\hbox{$\; \buildrel > \over \sim \;$}}
\newcommand\simlt{\lower.5ex\hbox{$\; \buildrel < \over \sim \;$}}
\newcommand\Sigdl{\Sigma_{\rm dl}}
\newcommand\aSigdl{\langle\Sigdl\rangle}
\newcommand\bSig{\bar\Sigma}

\shorttitle{Star Formation in Barred-spiral Galaxies}
\shortauthors{Seo \& Kim}

\begin{document}

\title{Effects of Spiral Arms on Star Formation in Nuclear Rings of Barred-spiral Galaxies}
\author{Woo-Young Seo and  Woong-Tae Kim} 
\affil{Center for the Exploration of the Origin of the Universe
(CEOU), Astronomy Program, Department of Physics \& Astronomy, Seoul
National University, Seoul 151-742, Republic of Korea}

\email{seowy@astro.snu.ac.kr,  wkim@astro.snu.ac.kr}
\slugcomment{Accepted for publication in the ApJ}

\begin{abstract}
We use hydrodynamic simulations to study the effect of spiral arms
on the star formation rate (SFR) occurring in nuclear rings of
barred-spiral galaxies. We find that spiral arms can be an efficient
means of gas transport from the outskirts to the central parts,
provided that the arms are rotating slower than the bar. While the
ring star formation in models with no-arm or corotating arms is
active only during about the bar growth phase, arm-driven gas
accretion makes the ring star formation both enhanced and prolonged
significantly in models with slow-rotating arms. The arm-enhanced
SFR is larger by a factor of $\sim3-20$ than in the no-arm model,
with larger values corresponding to stronger and slower arms.
Arm-induced mass inflows also make dust lanes stronger. Nuclear
rings in slow-arm models are $\sim45\%$ larger than in the no-arm
counterparts. Star clusters that form in a nuclear ring exhibit an
age gradient in the azimuthal direction only when the SFR is small,
whereas no noticeable age gradient is found in the radial direction
for models with arm-induced star formation.
\end{abstract}

\section{Introduction\label{sec:intro}}

Barred-spiral galaxies often host star-forming nuclear rings at
their centers (e.g.,
\citealt{phi96,but96,kna06,maz08,com09,san10,maz11,hsi11,van11}).
These rings are most likely produced by the radial infall of gas at
large radii caused by angular momentum loss due to its nonlinear
interactions with an underlying stellar bar potential (e.g,
\citealt{com85,shl90,ath92,hel94,kna95,but96,kim12,ks12}). They
appear smaller in more strongly barred galaxies (e.g.,
\citealt{com10}) and unrelated to resonances with the bars (e.g.,
\citealt{pin14}).  This suggests that the ring location is
determined primarily by the amount of angular momentum loss rather
than the resonances, as confirmed by numerical simulations (e.g.,
\citealt{ksk12,kim12}). Large gas surface density and small
dynamical timescale of nuclear rings make them one of the most
intense star forming regions in disk galaxies.

Observations indicate that the star formation rate (SFR) in a
nuclear ring varies widely from galaxy to galaxy (e.g.,
\citealt{maz08,com10}), although the total gas content in a ring is
similar at $\sim(1-6)\times 10^8\Msun$ (e.g.,
\citealt{but00,ben02,she05,sch06}). Data presented in \citet{maz08}
and \citet{com10} suggest that strongly barred galaxies tend to have
a small SFR, while weakly barred galaxies have a wide range of SFR
at  $\sim 0.1-10\Aunit$. Ring star formation appears to be long
lived, occurring either continuously \citep{van13} or episodically
\citep{all06,sar07} for $\sim1-3\Gyr$. In some galaxies, young star
clusters exhibit an azimuthal age gradient such that they tend to be
older farther away from the contact points between a ring and dust
lanes, while there is no noticeable age gradient in many other
galaxies (e.g, \citealt{bok08,maz08,ryd10,bra12}). Yet, what
determines the ring SFR as well as the presence or absence of the
azimuthal age gradient along a ring is not clearly understood.

In a recent attempt to understand these observational results,
\citet[hereafter Paper~I]{seo13} ran hydrodynamic simulations for
star formation occurring in nuclear rings of barred galaxies without
spiral arms, and found that the ring SFR is controlled mainly by the
mass inflow rate to the ring rather than the total gas mass in the
ring. In these bar-only models, the massive gas inflows caused by
the bar growth result in a strong burst phase of SFR that lasts only
for $\sim 0.2$ Gyr, after which both the mass inflow rate and SFR
are decreased to very small values below $\sim \,0.1\Aunit$. The
main reason for this short burst of star formation is that only the
gas inside the bar region (more precisely, inside the outermost
$x_1$ orbit) can respond to the bar potential to initiate the rapid
gas infall, while the gas outside the bar is not much affected by
the bar potential (e.g, \citealt{kim12,ks12}). Paper~I also found
that an azimuthal age gradient of young star clusters is expected
when the SFR is less than a critical value affordable at the contact
points. These bar-only models may explain ring star formation in
galaxies with low SFR, but would require that the bars should be
dynamically young in galaxies with high SFR, which is quite unlikely
since the lifetime of bars is quite long (several Gyrs) in
observations (e.g., \citealt{gad05,per09}) and $N$-body simulations
(e.g., \citealt{she04,bou05,ber07,ath13}).

If the mass inflow rate to the ring is really a critical factor in
determining the ring SFR, long-lived star formation requires that
fresh gas should be supplied to the nuclear rings continually or
continuously. There are several additional gas-feeding mechanisms at
work in real galaxies that may change the temporal evolution of the
SFR considerably. These include galactic fountains (e.g,
\citealt{fra06,fra08}), cosmic accretion of primordial gas (e.g,
\citealt{dek09,ric12}), and angular momentum dissipation by spiral
arms (e.g., \citealt{rob72,lub86,hop11,kim14}). For instance,
\citet{fra08} estimated the gas infalling rates of $\sim2.9$ and
$\sim0.8\Aunit$ for NGC 891 and NGC 2403, respectively, most of
which was ejected to the halos via supernova (SN) feedback, while
\citet{ric12} found high velocity clouds feed a normal galactic disk
with gas at a rate of $\sim0.7\Aunit$. Since the gas supply in the
form of fountains and cosmic accretion occurs over a whole disk
plane, the accreted gas should make its way to the galaxy center
anyway to help promote star formation in the rings.  Very recently,
\citet{kim14} showed that stellar spiral arms can play such a role,
transporting the gas inward at a rate $\sim 0.05-3.0\Aunit$
depending on the arm strength and pattern speed, which can
potentially enhance the ring SFR.

In this paper, we investigate star formation in a nuclear ring of a
disk galaxy that possesses both spiral arms and a bar.  This is a
straightforward extension of Paper~I that considered only the bar
potential. By varying the arm strength and pattern speed while
fixing the bar parameters, we quantity the effect of the spiral arms
on the ring SFR and gaseous structures that form. In Section
\ref{sec:modmeth}, we describe our galaxy models and numerical
method. In Section \ref{sec:res}, we present the results on overall
evolution of our galaxy models, star formation occurring in nuclear
rings, and age gradients of star clusters.  We summarize and discuss
our results in Section \ref{sec:sum}.

\renewcommand{\arraystretch}{1.6}
\begin{deluxetable*}{ccccccccc}
\tabletypesize{\footnotesize} \tablewidth{0pt} \tablecaption{Model
Parameters and Simulation Outcomes \label{tbl:model}} \tablehead{
\colhead{Model} & \colhead{$\Farm$}& \colhead{$\Oms$} &
\colhead{$\DSN $} & \colhead{$\aSigdl$} & \colhead{$\aMring$} &
\colhead{$M_*$} & \colhead{$\aSFR$} &
\colhead{$\Rring$} \\
\colhead{} & \colhead{(\%)}& \colhead{($\rm km\;s^{-1}\kpc^{-1})$} &
\colhead{$(\rm Myr)$} & \colhead{$(\rm M_\odot \pc^{-2})$} &
\colhead{$(10^9 \Msun)$} & \colhead{$(10^9 \Msun)$} & \colhead{$(\rm
M_\odot \,yr^{-1})$} &
\colhead{$(\rm kpc)$} \\
\colhead{(1)} & \colhead{(2)} & \colhead{(3)} & \colhead{(4)} &
\colhead{(5)} & \colhead{(6)} & \colhead{(7)} & \colhead{(8)} &
\colhead{(9)} }
 \startdata
 F00    & 0 & -  & 10 & 21.8$\pm$9.34   & 0.25$\pm$0.01 & 0.92 & 0.19$\pm$0.21 & 0.67 \\
  \hline
 F05P10 & 5 & 10 & 10 & 41.3$\pm$34.6   & 0.21$\pm$0.06 & 1.53 & 0.98$\pm$0.76 & 0.90 \\
 F05P20 & 5 & 20 & 10 & 34.8$\pm$31.2   & 0.23$\pm$0.03 & 1.27 & 0.62$\pm$0.35 & 0.89 \\
 F05P33 & 5 & 33 & 10 & 23.1$\pm$13.3   & 0.27$\pm$0.01 & 1.02 & 0.31$\pm$0.20 & 0.72 \\
 \hline
 F10P10 & 10& 10 & 10 & 66.7$\pm$55.5   & 0.26$\pm$0.08 & 2.07 & 1.32$\pm$1.09 & 1.00 \\
 F10P20 & 10& 20 & 10 & 75.4$\pm$79.2   & 0.27$\pm$0.05 & 1.70 & 1.32$\pm$1.07 & 0.95 \\
 F10P33 & 10& 33 & 10 & 24.0$\pm$20.5   & 0.28$\pm$0.01 & 1.01 & 0.31$\pm$0.17 & 0.73 \\
 \hline
 F15P10 & 15& 10 & 10 & 101.7$\pm$133.8 & 0.24$\pm$0.07 & 2.84 & 2.44$\pm$1.93 & 0.98 \\
 F15P20 & 15& 20 & 10 & 94.2$\pm$90.1   & 0.25$\pm$0.07 & 2.31 & 2.11$\pm$1.41 & 1.02 \\
 F15P33 & 15& 33 & 10 & 24.4$\pm$19.1   & 0.26$\pm$0.01 & 1.16 & 0.34$\pm$0.40 & 0.70 \\
 \hline
 F20P10 & 20& 10 & 10 & 123.9$\pm$314.1 & 0.23$\pm$0.08 & 3.55 & 3.54$\pm$1.69 & 1.08 \\
 F20P20 & 20& 20 & 10 & 114.3$\pm$220.3 & 0.27$\pm$0.07 & 3.04 & 2.95$\pm$1.84 & 0.94 \\
 F20P33 & 20& 33 & 10 & 30.1$\pm$27.7   & 0.23$\pm$0.03 & 1.42 & 0.75$\pm$0.42 & 0.81 \\
 \hline
 F10P20d &10& 20 &  5 & 68.3$\pm$80.4   & 0.25$\pm$0.06 & 1.76 & 1.45$\pm$1.21 & 0.95
\enddata
\tablecomments{$\Farm$ denotes the dimensionless arm strength;
$\Oms$ is the arm pattern speed; $\DSN$ is the time interval between
the cluster formation and SN explosion; $\aSigdl$ is the mean
surface density of the dust-lane segments at $R=2.0-2.5\kpc$
averaged over $t=0.4-1.0\Gyr$; $\aMring$ is the gas mass in the ring
averaged over $t=0.4-1.0\Gyr$; $M_*$ is the total stellar mass
formed until $t=1.0\Gyr$; $\aSFR$ is the ring SFR averaged over
$t=0.4-1.0\Gyr$; $\Rring$ is the ring radius at $t=1.0\Gyr$.}
\end{deluxetable*}

\section{Model and Method}\label{sec:modmeth}

We consider disk galaxies with both spiral arms and a bar.  Our
galaxy models are identical to those in Paper~I except that we
initially consider an exponential gaseous disk rather than a uniform
disk and that we additionally include stellar spiral perturbations.
The reader is referred to Paper~I for the detailed description of
the simulation setups and numerical methods. Here we briefly
describe our current models and methods.

The gaseous disk is infinitesimally-thin, self-gravitating,
unmagnetized, and rotating about the galaxy center. The initial
profile of gas surface density is taken to
\begin{equation}
\Sigma_0=29.4 \exp(-R/9.7\kpc)\Surf,
\end{equation}
which describes nearby disk galaxies reasonably well \citep{big12}.
We adopt an isothermal equation of state with sound speed of
$c_s=10\kms$.

The axisymmetric part of the external gravitational potential gives
rise to a rotational velocity profile that resembles normal disk
galaxies with velocity of $v_c\simeq 200\kms$ at the flat part.  The
non-axisymmetric part consists of two components: a bar and spiral
arms. As in Paper~I, the bar potential is modeled by a Ferrers
prolate spheroid whose parameters are fixed to the central density
concentration index  $n=1$, the semi-major and minor axes  $5\kpc$
and $2\kpc$, respectively, the mass $1.5\times10^{10}\Msun$, and the
pattern speed $\Omb = 33\kms\kpc^{-1}$. For the spiral potential, we
take a two-armed trailing logarithmic model of \citet{she06}:
\begin{equation}\label{eq:sp1}
\Phi_{s}(R,\phi;t)=\Phi_{s0}\cos\left( m\left[\phi+\frac{\ln R}{\tan
p_*} -\Oms t+\phi_0 \right] \right),
\end{equation}
for $R\geq6\kpc$ and $\Phi_{s}=0$ at $R<5\kpc$, with $\Phi_s$ between $5\kpc$ and
$6\kpc$ tapered by a Gaussian function. Here, $m$, $p_*$, $\Oms$,
and $\phi_0$ denote the number, the pitch angle, the pattern speed,
and the initial phase of the arms, respectively. The amplitude
$\Phi_{s0}$ of the arm potential is controlled by the dimensionless
arm-strength parameter $\Farm$ defined by
\begin{equation}\label{eq:sp2}
\Farm \equiv \frac{m\Phi_{s0}}{v_c^2 \tan p_*},
\end{equation}
which measures the radial force due to the spiral arms relative to
the centrifugal force from the background galaxy rotation (e.g.,
\citealt{kim14}). In this work, we fix $m=2$, $p_*=20^{\circ}$,
$\phi_0=147\degr$, and vary $\Farm$ and $\Oms$.

Our calculations incorporate a prescription for star formation and
ensuing feedback via SNe. We determine star-forming regions based on
the critical density corresponding to the Jeans condition, and allow
for a star formation efficiency of 1\% (e.g.,
\citealt{kru05,kru07}). When a cloud is determined to undergo star
formation, we spawn a sink particle corresponding to a star cluster,
and convert 90\% of the gas mass to the particle. The mass of each
particle is typically in the range of $\sim 10^5-10^7\Msun$. Each
particle interacts gravitationally with each other while orbiting
under the influence of total gravity, and injects radial momentum to
the surrounding gaseous medium, mimicking multiple simultaneous SN
explosions from a cluster. We consider a time delay, $\DSN$, between
star formation and SN feedback. Under the \citet{kro01} initial mass
function, the mass-weighted mean main-sequence lifetime of stars
with $M\geq 8\Msun$ that explode as SNe is estimated to be
$\DSN=10\Myr$, but we also run a case with $\DSN=5\Myr$ to study its
effect on the ring SFR. In our models, the amount of the radial
momentum per single SN in the in-plane direction is taken to be
$2.25 \times10^5\Msun\kms$. This corresponds to the snow-plow phase
of a shell expansion due to an injection of SN energy $10^{51}\erg$
in the in-plane direction (e.g., \citealt{che74, cio88, tho98,
kok13, kimm14}).

Although it is challenging to measure the pattern speeds of bars and
spiral arms, observations indicate that they are either corotating
or the arms are rotating more slowly than the bar (e.g.,
\citealt{fat09,mar11}). To explore various situations, we run a
total of 14 models with differing $\Farm$ between 0 and 20\%, $\Oms$
between 10 and $33\freq$, and $\DSN$ between 5 and $10\Myr$. Columns
(1)--(4) of Table \ref{tbl:model} list the name and the parameters
of each model. Columns (5)--(9) give some of the simulation
outcomes, which will be explained later. Model F00 is a bar-only
model, while the other models possess spiral arms as well. Model
F10P20d is a control model with $\DSN=5\Myr$. Note that the models
with $\Oms=33\freq$ have the arms and bar corotating, with the
co-rotation resonance (CR) radius at $\RCRb=\RCRa=6\kpc$, while the
models with $\Oms=10$ and $20\freq$ have the $\RCRa=20$ and
$10\kpc$, respectively.

As in Paper~I, we integrate the basic ideal hydrodynamic equations
using the CMHOG code  in the frame corotating with the bar
\citep{pin95}. To resolve the ring regions with high accuracy, we
set up a logarithmically-spaced grid that extends from $R = 0.05$ to
$30\kpc$. The number of zones in our models is $1290\times 632$ in
the radial and azimuthal directions covering the half-plane with
$\phi = -\pi/2$ to $\pi/2$, leading to the grid size of $5\pc$ at $R
= 1 \kpc$ where a ring preferentially forms. We adopt the outflow
and periodic boundary conditions at the radial and azimuthal
boundaries, respectively. In order to avoid strong transients in the
gas flow caused by a sudden introduction of the bar, the bar is
slowly introduced over one bar revolution time of 0.19 Gyr.

\section{Simulation Results}\label{sec:res}

We in this section first describe the overall evolution of our
fiducial model F10P20 with $\Farm=10\%$ and $\Oms=20\freq$ in
comparison with its no-arm counterpart, Model F00. Evolution of
other models with arms is quantitatively similar to that of Model
F10P20. We then present the results on star formation histories and
distributions of star clusters that form in nuclear rings.

\begin{figure*}
\epsscale{1.17} \plotone{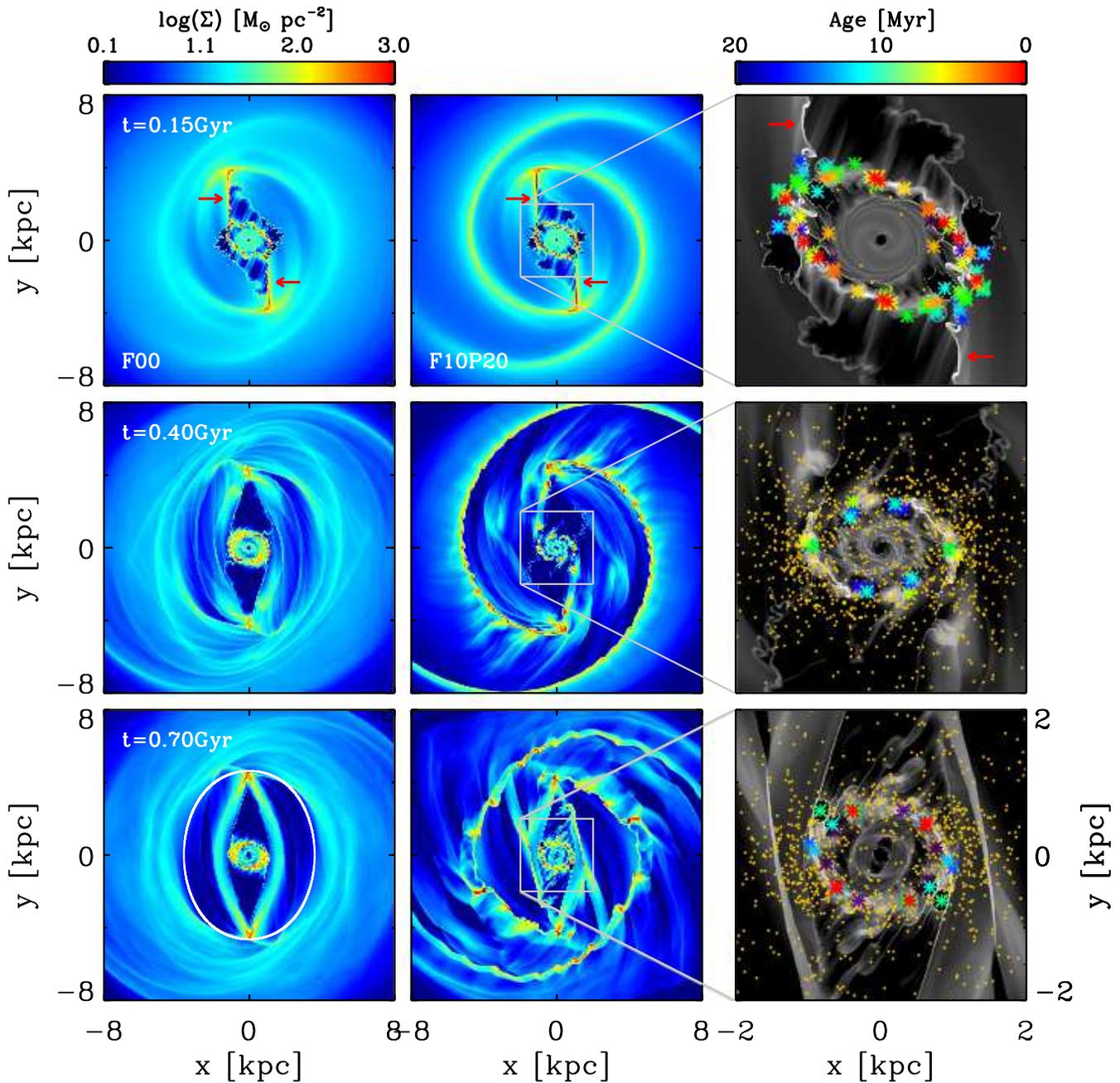} \caption{Snapshots of gas surface
density at $t=0.15\Gyr$ (top), $0.40\Gyr$ (middle), and $0.70\Gyr$
(bottom) for Model F00 (left column) and Model F10P20 (middle
column). The right panels expand the central $2\kpc$ regions of
Model F10P20 to show the distributions of young clusters with age
less than $20\Myr$ (asterisks) and older clusters (dots) that
formed. The CR of the bar is at $\RCRb=6\kpc$ inside which the gas
is rotating in the counterclockwise direction. A pair of the red
arrows in each of top panels indicate high-density ridges, termed
dust lanes, located at the downstream side from the bar major axis.
The oval in the lower-left panel draws the outermost $x_1$ orbit,
outside of which the gas distribution is not much perturbed in Model
F00. The left and right colorbars label $\log \Sigma$ and the age of
young clusters, respectively. \label{fig:all}}
\end{figure*}

\subsection{Overall Evolution}\label{sec:over}

Figure \ref{fig:all} shows snapshots of the gaseous surface density
in logarithmic scale at $t = 0.15$, 0.4, and $0.7\Gyr$ of Models F00
with no arms (left) and  Model F10P20 with arms (middle). The right
panels zoom in the central $2\kpc$ regions of Model F10P20 to
display the positions of star-forming regions younger than $20\Myr$
(colored asterisks) as well as all star clusters that have formed
(small dots). The bar is pointing toward the $y$-axis and remains
stationary in the simulation domain. The solid oval in the
lower-left panel draws the outermost $x_1$ orbit under the given
potential, which cuts the $x$- and $y$-axes at $3.6\kpc$ and
$4.7\kpc$, respectively. The total gas mass enclosed by the
outermost $x_1$ orbit is $1.2\times 10^9\Msun$ in the initial disk.
Note that the inner ends of the gaseous spiral arms are connected to
the bar ends for most of the time, even though the arms and the bar
have different pattern speeds.

An introduction of the bar and spiral potentials provides strong
perturbations for gas orbits that would otherwise remain circular.
The gas in the bar region is readily shocked to form ``dust lanes''
referring to high-density ridges, indicated as arrows in the upper
panels of Figure \ref{fig:all}, located at the leading side of the
bar major axis (e.g., \citealt{ath92}). Gas passes through the
dust-lane shocks almost perpendicularly and loses angular momentum,
moving radially inward to form a nuclear ring at the location where
centrifugal force balances the gravity \citep{ksk12,kim12}. At the
same time, the gas in the arm region develops spiral shocks whose
pitch angle is smaller than that of the stellar arms.  The offset
between the pitch angles of gaseous and stellar arms is larger for
smaller $\Farm$ and larger $\Oms$ \citep{kim14}. In general, spiral
shocks in the arm region are much weaker than the dust-lane shocks
in the bar region owing largely to a smaller angle between the gas
streamlines and the shock fronts in the former, so that gas infall
due to the spiral shocks occurs much more slowly than that
associated with the bar. At about $0.11 \Gyr$, stars start to form
in the ring where plenty of gas is accumulated by the bar potential
to meet the Jeans condition for gravitational collapse.

At early time ($t=0.15\Gyr$), the effect of spiral arms on the bar
region is almost negligible since they are still weak and growing.
When the arms become strong enough to induce spiral shocks, the gas
originally located outside the bar region but inside the CR of the
arms (i.e., $\RCRb < R < \RCRa$) starts to move radially inward by
losing angular momentum due to the spiral potential and associated
spiral shocks, while the gas outside $\RCRa$ drifts outward. The
inflowing gas due to the arms moves on along $\xone$ orbits after
entering the bar region. The gas is piled up at the bar ends where
$\xone$ orbits crowd. Mutual collisions of gas orbits there further
take away angular momentum from the gas, intermittently sending gas
blobs along the dust lanes to the nuclear ring.  When this happens,
the dust lanes become inhomogeneous, as illustrated in the
$t=0.4\Gyr$ snapshot of Model F10P20 in Figure \ref{fig:all}. This
not only enhances the gas surface density of the dust lanes but also
fuels episodic star formation in the ring at late time (see below).

\begin{figure}
\epsscale{1.2} \plotone{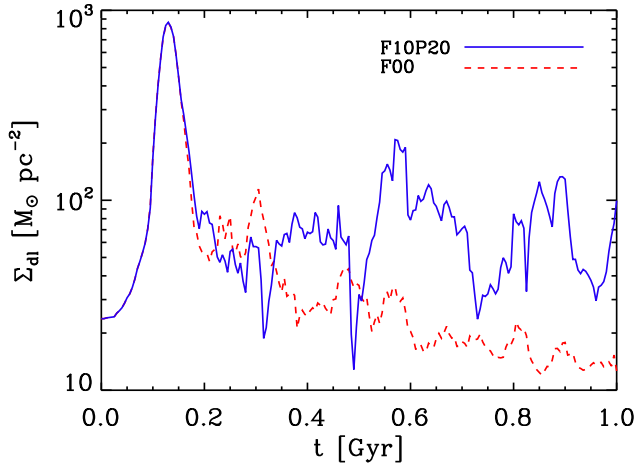} \caption{Temporal changes of the
gas surface density of the dust lanes $\Sigdl$ averaged over
$R=2.0-2.5\kpc$ for Models F10P20 and F00. In Model F00, the dust
lanes are strong only when the gas in the bar region infalls ($t
\sim 0.1-0.2\Gyr$) and when the ring gas is expelled to the bar
region by star formation feedback ($t\sim 0.2-0.3\Gyr$). The
arm-induced mass inflows make the dust lane strong at late time in
Model F10P20. \label{fig:dl}}
\end{figure}

Dust lanes are located at the downstream side of galaxy rotation
from the bar major axis, and are more straight in more strongly
barred galaxies \citep{kna02,com09,ksk12}. In our models, they
typically have a width of $\sim50\pc$ and extend from the bar ends
to the nuclear ring located at $R\sim 1\kpc$.  Figure \ref{fig:dl}
plots temporal variations of the mean gas surface density, $\Sigdl$,
of the dust-lane segments at $R = 2.0-2.5\kpc$ in Models F00
(dashed) and F10P20 (solid). In Model F00, $\Sigdl$ is large only
when the gas in the bar region experiences the massive infall to the
nuclear ring ($t\sim0.1- 0.2\Gyr$) and when the feedback from the
initial starburst activities sends the ring material out to the dust
lanes ($t\sim0.2-0.3\Gyr$), after which $\Sigdl$ decays to very
small values. The early behavior of $\Sigdl$ in Model F10P20 is
similar to that in Model F00, but the gas inflows induced by spiral
arms make the dust lanes in the former much more pronounced at
$t\simgt 0.4\Gyr$ than in the no-arm counterpart. This trend can
also be seen in Figure \ref{fig:all} where the dust lanes in Model
F00 are strong at $t=0.15\Gyr$ but can be barely identified at
$t\simgt 0.4\Gyr$, while they are vividly apparent at any time in
Model F10P20.  Column (5) of Table \ref{tbl:model} gives the mean
surface density $\aSigdl$ as well as standard deviations of the dust
lanes for all models, where the angle brackets $\langle\,\rangle$
represent a time average over $t=0.4-1.0\Gyr$.

In the $t=0.7\Gyr$ snapshot of Model F00, there is a well-defined,
elongated gaseous feature, called a gaseous inner ring, that lies
inside the outermost $x_1$ orbit and encompasses the dust lanes
(e.g., \citealt{but86,but13,reg02}). The inner ring has roughly the
same size as the bar (e.g., \citealt{but96}). As explained in
Paper~I, it begins to form after dust lanes find their equilibrium
positions by collecting the residual gas that did not experience the
dust-lane shocks. In Model F10P20, however, the inner ring is
strongly perturbed by the arm-induced mass inflows at late time. It
is also influenced by feedback from star formation in the nuclear
ring. As Figure \ref{fig:all} shows, the spiral shocks at
$t=0.4\Gyr$ are abundant with dense clumps produced by a wiggle
instability of the shock fronts, which occurs as a consequence of
the potential vorticity accumulation in the gas flows moving across
curved shock fronts multiple times \citep{kkk14}. These clumps
collide and merge with each other as they move along the arms, and
become loose at the interface between the arm and bar regions. When
they enter the bar region, they thus have significant inward radial
velocities, providing strong perturbations to the gas already in the
inner ring. They eventually settle on $x_1$ orbits, lose angular
momentum by hitting the bar ends, and move further in to the nuclear
ring.

\subsection{Star Formation}\label{sec:sf}

\begin{figure}
\epsscale{1.23} \plotone{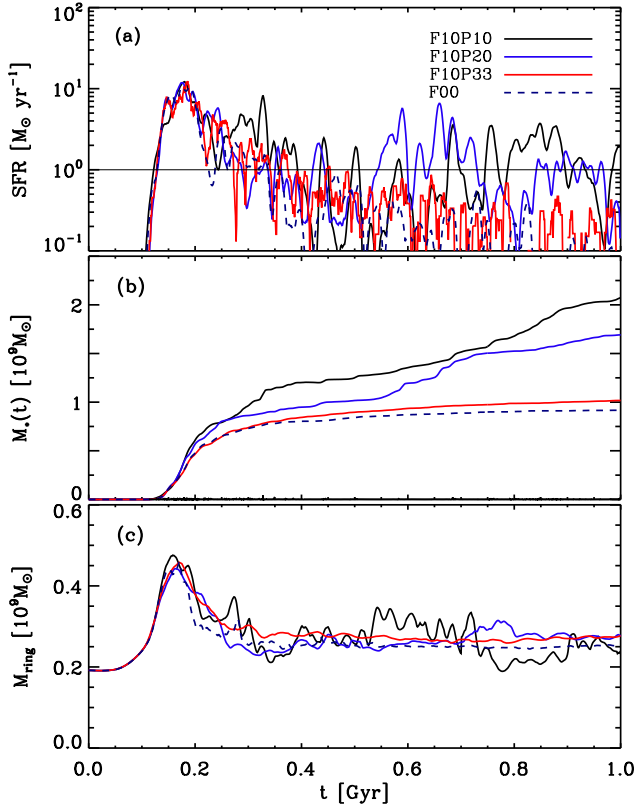} \caption{Temporal variations of
the (a) SFR in the ring, (b) total stellar mass $M_*(t)$ formed
until $t$, and (c) total gas mass $\Mring$ in the ring for models
with $\Farm=10\%$ and varying $\Oms$. The results of the bar-only
model are compared as dashed lines. Arm-induced gas inflows make the
SFR rejuvenated at $t\simgt0.4\Gyr$. The ring SFR is not well
correlated with $\Mring$. \label{fig:SFR}}
\end{figure}

\subsubsection{Enhanced SFR by Arms}\label{sec:sfr}

Figure \ref{fig:SFR} plots the temporal evolution of the SFR in the
ring, the total stellar mass $M_*(t)$ formed until time $t$, and the
total gas mass $\Mring$ in the ring for models with $\Farm=10\%$ but
differing $\Oms$.\footnote{We calculate the ring mass as
$\Mring\equiv \int_{0.5\kpc}^{1.5\kpc}\int \Sigma R d\phi dR$, since
the ring density in our models is about two orders of magnitude
larger than the density of the surrounding medium near the galaxy
center.} The results of Model F00 with $\Farm=0$ are compared as
dashed lines. The overall behavior of the SFR in the bar-only model
is characterized by a strong primary burst and a few subsequent
secondary bursts before declining to small values at $t\simgt
0.3\Gyr$, with the duration of the burst phase corresponding to the
bar growth time (Paper~I). The presence of spiral arms especially
when the pattern speed is small can make the SFR rejuvenated at
$t\simgt 0.4\Gyr$. As we mentioned above, the mass infalls from the
bar ends to the ring occur intermittently, resulting in episodic
star formation in the ring at late time. Since the typical inflow
velocity due to the arms is $\sim1\kms$ \citep{kim14}, the
arm-induced SFR in the ring can persist longer than the Hubble time,
as long as $\RCRa$ is located sufficiently far away from the bar
ends. Note that ring SFR is not enhanced much in Model F10P33, since
$\RCRa$ is located just outside the bar ends.

\begin{figure}
\epsscale{1.23} \plotone{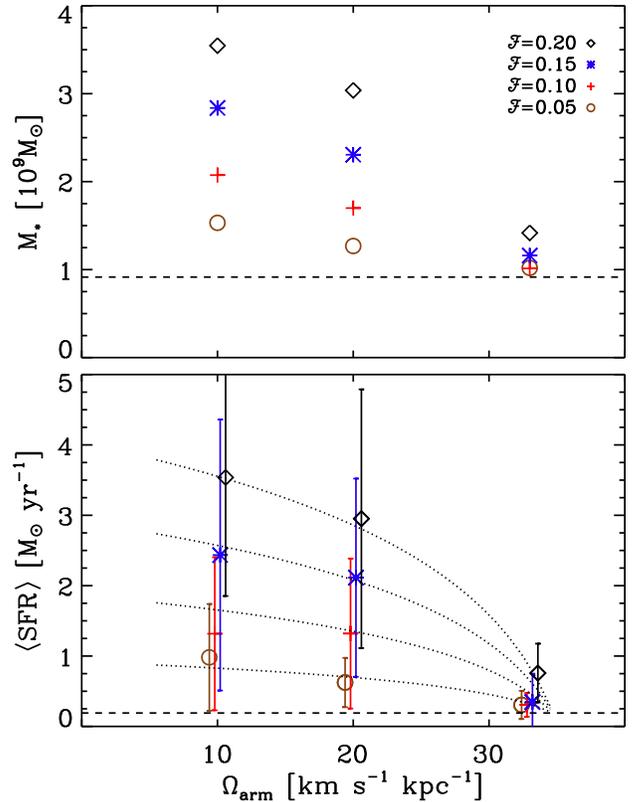} \caption{(a) Total stellar mass
$M_*$ formed until $t=1\Gyr$ and (b) the mean values (symbols) and
standard deviations (errorbars) of the SFR averaged over
$t=0.4-1\Gyr$ as functions of the arm strength $\Farm$ and pattern
speed $\Oms$. The horizontal dashed line in each panel marks the
value in the bar-only model. In (b), the data are displaced slightly
in the horizontal direction for clarity.  The dotted lines are our
best fits (Eq.\ (\ref{e:fits})). \label{fig:Mstar}}
\end{figure}

Figure \ref{fig:SFR}(c) shows that in all models $\Mring$ is
maintained relatively constant at $\sim(2-4)\times10^8\Msun$
throughout evolution, similar to observations (e.g.,
\citealt{she05}).  Column (6) of Table \ref{tbl:model} gives
$\aMring$ for all models.  Since the SFR depends on $\Oms$
considerably, this suggests that it is not the gas mass in the ring
but the mass inflow rate to the ring that controls the ring SFR,
even when the effect of spiral arms is included. With the typical
ring radius of $1\kpc$ and thickness of $50\pc$, this translates
into the averaged gas surface density $\Sigma_{\rm ring} \sim
650-1250 \Surf$, which is approximately equal to the critical
density $\Sigma_c = 10^3 \Surf$ for Jeans collapse. When
$\Sigma_{\rm ring}>\Sigma_c$, star formation takes place in the
ring, reducing the gas density. When $\Sigma_{\rm ring}<\Sigma_c$,
star formation is halted in the ring which has to wait until fresh
gas is filled in to resume star formation. For the bar-only models,
Paper~I found that the ring shrinks in size steadily over time due
to the addition of gas with low angular momentum.  Enhanced SFR by
arms turns out to reduce the shrinking rate of the ring size (see
Section \ref{sec:age}).

Figure \ref{fig:Mstar} plots $M_*$ at $t=1\Gyr$ and the mean values
$\aSFR$ (symbols) along with the standard deviations (errorbars) of
the SFR averaged over $t=0.4-1.0\Gyr$ for all models.  These values
are also tabulated in Columns (7) and (8) of Table \ref{tbl:model}.
The horizontal dashed line in each panel represents the case with no
arm.  Note that $M_*=0.9\times 10^9\Msun$ in Model F00, which is
about 75\% of the initial gas mass inside the outermost $x_1$ orbit.
While $M_*/(1\Gyr)$ corresponds to the mean SFR throughout entire
evolution, $\aSFR$ measures the mean value of the arm-induced SFR
after the initial gas infall due to the bar potential is almost
finished. Clearly, both $M_*$ and $\aSFR$ are larger for models with
larger $\Farm$ and/or smaller $\Oms$. For $\Oms=10$ and $20\freq$,
for example, the time-averaged ring SFR over $t=0.4-1.0\Gyr$ is
enhanced by a factor of 5.2, 6.9, 12.8, 18.6, and 3.3, 6.9, 11.1,
15.5 for models with $\Farm=5, 10, 15, 20\%$, respectively, compared
to the no-arm counterpart.  The dotted lines are our best fits to
$\aSFR$:
\begin{equation}\label{e:fits}
 \frac{\aSFR} {\rm M_\odot\;yr^{-1}} =
 0.19 + 25 \Farm^{1.2}  \log \left( 11.5- 10
 \frac{\Oms}{\Omb}\right).
\end{equation}
The increasing trend of the ring SFR with $\Farm$ is similar to
that of the gas inflow rates driven by the arms reported by
\citet{kim14}. This makes sense since a smaller pattern speed
implies a larger $\RCRa$ and since stronger spiral shocks can remove
a larger amount of angular momentum from the gas. When the arms are
corotating with the bar, on the other hand, the presence of spiral
arms does not affect the ring SFR much. This shows that the
enhancement of the ring SFR due to spiral arms is significant only
when they have different pattern speeds.

\subsubsection{Age Gradients}\label{sec:age}

Paper~I showed that young star clusters exhibit a noticeable age
gradient in the azimuthal direction along a nuclear ring only when
the SFR is larger than the critical value $\Mcrit$, which is set by
the maximum SFR affordable at the contact points. In the high-SFR
phase ($\SFR > \Mcrit$), star-forming regions are distributed
randomly throughout the ring, hence no age gradient of clusters is
expected. In the low-SFR phase ($\SFR<\Mcrit$), on the other hand,
they are localized to the contact points, leading to a well-defined
azimuthal age gradient. These two modes of star formation are
referred to respectively as ``popcorn" and ``pearls-on-a-string"
models by \citet{bok08}. For our adopted parameters, Paper~I found
$\Mcrit\sim1\Aunit$, although it depends sensitively on the ring
size and the gas sound speed. Paper~I also showed that star clusters
with age $<1\Gyr$ show a positive radial age gradient such that
older clusters are located at larger $R$ owing to a secular decrease
of the ring size.

We find that the condition for the presence or absence of the
azimuthal age gradient of young clusters is not affected by
arm-enhanced star formation. By analyzing the dependence of the
cluster ages younger than $10\Myr$ on the azimuthal positions in all
models with spirals, we find that star formation occurs in the
pearls-on-a-string fashion for $\sim80\%$ of the low-SFR phase and
for $\sim20\%$ of the high-SFR phase. In the high-SFR phase, the
mass inflow rate along the dust lanes is too large for star
formation to consume all the inflowing gas at the contact points:
overflowing gas produces star-forming clumps distributed randomly
along the nuclear ring.  In the low SFR phase, however, most of the
inflowing gas undergoes star formation at the contact points.  Since
clusters age as they move along the ring, this naturally leads to an
azimuthal age gradient.

\begin{figure}
\epsscale{1.20} \plotone{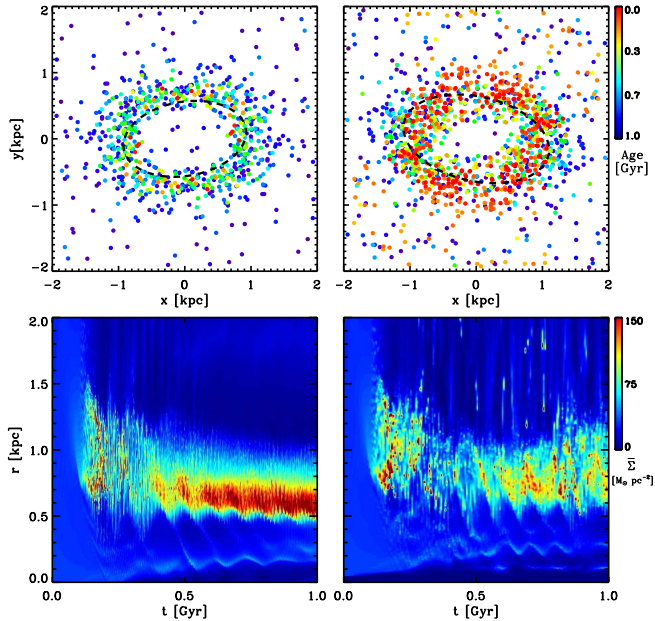} \caption{(Upper) Spatial
distributions of star clusters at $t=1\Gyr$, with color denoting
their age and (lower) the radial and temporal variations of the
azimuthally-averaged surface density $\bSig$. The left and right
panels are for Model F00 and F20P20, respectively. In the upper
panels, the dashed oval fits the ring at $t=1\Gyr$ in each model.
 \label{fig:grad}}
\end{figure}

However, the arm-enhanced star formation tends to remove the radial
age gradient of star clusters. The upper panels of Figure
\ref{fig:grad} display the spatial distributions of star clusters at
$t=1\Gyr$ in Model F00 (left) and Model F20P20 (right), with the
color representing their age. The lower panels plot the
corresponding temporal changes of the azimuthally-averaged gas
surface density $\bSig(R)\equiv \int\Sigma d\phi/(2\pi)$ in linear
scale. In the no-arm model, younger clusters are located
preferentially at smaller $R$ due to a secular decrease in the ring
size. This results in a radial age gradient amounting to $d
\log(t/{\rm yr})/d(R/{\rm kpc}) \sim 15$ in Model F00. In Model
F20P20, on the other hand, gas driven in by the spiral arms has
larger angular momentum than in the ring gas, so that the ring does
not decrease in size after $\sim0.4\Gyr$.  In addition, active star
formation feedback disperses the ring gas widely in the radial
direction, making the ring larger than in the no-arm model.  Column
(9) of Table \ref{tbl:model} gives the average ring radius $\Rring$
at $t=1\Gyr$, where $\Rring\equiv \int R \bSig dR/\int \bSig dR$,
with the radial integration taken over $R=0.5-1.5\kpc$, showing that
the radii of the rings in spiral-arm models with $\Oms\simlt20\freq$
are larger by about 45\% than in Model F00. Consequently, star
clusters in models with arm-enhanced star formation exhibit no
apparent radial age gradient, as the upper-right panel of Figure
\ref{fig:grad} illustrates.

\subsubsection{Effects of $\DSN$}

\begin{figure}
\epsscale{1.2} \plotone{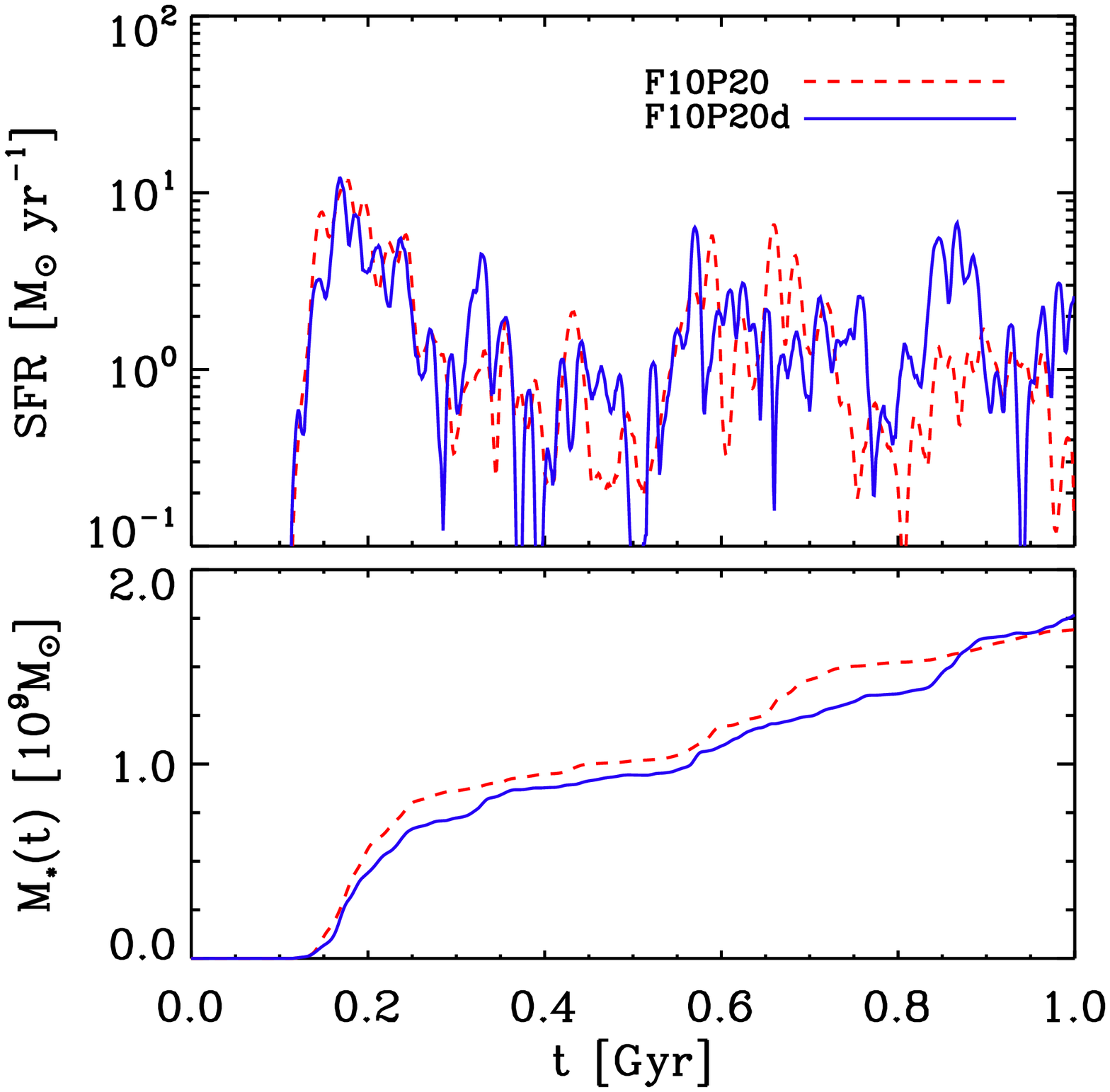} \caption{Comparison of the ring
SFR and the total stellar mass $M_*(t)$ between Model F10P20 with
$\DSN=10\Myr$ and Model F10P20d with $\DSN=5\Myr$. While the
detailed histories of the SFR are different, the time-averaged SFR
and $M_*$ are within 10\%.\label{fig:tau}}
\end{figure}

To explore the effects of the time delay between star formation and
SN explosion, we run Model F10P20d with $\DSN=5\Myr$, while the
other parameters are taken identical to those in Model F10P20.
Figure \ref{fig:tau} compares the temporal changes of the SFR and
$M_*(t)$ from these two models.  Due to a shorter delay, feedback
occurs earlier in Model F10P20d, which tends to reduce SFR and $M_*$
at early time compared to those in Model F10P20. Although detailed
star formation histories are different, the time averaged SFRs over
$t= 0.4-1.0 \Gyr$ are $1.32$ and $1.45\Aunit$ for Models F10P20 and
F10P20d, respectively, which agree within 10\%. The total stellar
mass formed at the end of the runs are also similar at $M_*=1.70
\times 10^9\Msun$ and $1.76 \times 10^9\Msun$ for Models F10P20 and
F10P20d, respectively. This demonstrates that the results on the
arm-enhanced SFR presented in this work is insensitive to the choice
of $\DSN$ as long as it is within a reasonable range.

\section{Summary and Discussion}\label{sec:sum}

We have presented the results of two-dimensional grid-based
hydrodynamic simulations to study star formation in nuclear rings of
barred-spiral galaxies. The gaseous medium is taken to be
isothermal, self-gravitating, unmagnetized, and limited to an
infinitesimally-thin disk.  We incorporate a prescription for star
formation and delayed feedback via SNe in the form of momentum
injection. We handle the bar and spiral patterns using
rigidly-rotating, fixed gravitational potentials, and do not
consider the back reaction of the gas to the underlying patterns. To
study various situations, we vary only the strength and pattern
speed of the arms, while fixing other arm and bar parameters. The
main results and corresponding discussions are as follows.

1. -- \emph{Arm-enhanced SFR}: Spiral arms located in the outer
disks can drive gas toward the bar region, enhancing star formation
in nuclear rings considerably, only if the arm pattern speed is
smaller than that of the bar. This is because only the gas located
between the bar ends and the CR of the arms can lose angular
momentum by passing through spiral arms and move inward to the bar
region, while the gas outside the CR of the arms move radially
outward. The gas entering the bar region is first gathered to the
bar ends where it loses its angular momentum additionally to move
further in along the dust-lane shocks to the nuclear ring. The
inflow to the ring occurs in an intermittent fashion, making the
ring star formation episodic that lasts long until the end of the
simulations. Ignoring the star formation induced at early time by
the bar, the enhanced SFR by spiral arms is $\sim(0.6-3.5)\Aunit$,
which is about 3 to 20 times larger than that in the no-arm
counterpart. This has an important implication that the SFR
histories in nuclear rings can be significantly affected by spiral
arms that are slow and strong. Some galaxies are observed to have
undergone several episodic bursts of star formation in the rings
over a few Gyrs \citep{all06,sar07,van13}, which appear difficult to
be explained by gas inflows driven solely by the bar potential
(Paper~I). We suggest that these sustained starburst activities
might have resulted from additional gas feeding due to spiral arms.

2. -- \emph{Dust Lane Strength}: The mass inflows driven by arms
also help make the dust lanes stronger. By analyzing Sloan Digital
Sky Survey DR7 data, \citet{com09} found that about 20\% of 266
galaxies with measured bar strength host dust lanes with appreciable
strength. On the other hand, numerical simulations with only a bar
potential show that dust lanes remain strong only for $\sim0.2\Gyr$
around the time when the bar potential achieves its full strength
(e.g, Paper~I; \citealt{ks12}). Given that bars persist for several
Gyrs (e.g., \citealt{gad05,per09,ber07,ath13}), this is not
compatible with the results of \citet{com09}.  Barred galaxies with
strong dust lanes may be either dynamically young or supplied with
fresh gas. Our numerical results in this paper suggest that spiral
arms with large $\RCRa$ can be efficient to transport the gas from
outside to the bar region. The typical density of dust lanes in our
standard model F10P20 is about $70\Surf$. This corresponds to the
extinction magnitude of $A_V \sim 5$ in the visual band assuming a
standard value of $\sim 3$ for the ratio of total to selective
extinction, readily visible against the background stellar light.

3. -- \emph{Age Gradients of Star Clusters}: The gas driven in from
the arms to the ring is found to have larger angular momentum than
the gas already in the ring. In addition, feedback from active star
formation at late time tends to reduce the rate of angular momentum
removal, making the rings in models with spiral arms larger by
$\sim45\%$ than those in bar-only models. Consequently, star
clusters formed in bar-only models retain an age gradient in the
radial direction, while they do not in models with slow-rotating
spiral arms. On the other hand, the arm-enhanced star formation
exhibits an azimuthal age gradient of star clusters such that
younger clusters are located closer to the contact points when SFR
is small, while clusters with different ages are well mixed when SFR
is large.  The critical SFR that determines the absence/presence of
the azimuthal age gradient is set by the maximum gas consumption
rate $\Mcrit$ at the contact points, which is $\sim 1\Aunit$ in our
current models, although it may depends sensitively on the gas sound
speed and the ring size as $\Mcrit \propto c_s^3\Rring^2$ (Paper~I;
see also \citealt{ksk14}). This appears consistent with the
observational data of \citet{maz08} in that nuclear rings with an
azimuthal age gradient have, on average, a smaller SFR than those
without age gradient.

Many uncertainties surround the observational determinations of the
SFR in nuclear rings and the arm pattern speed. The derived SFRs
from different methods do not always agree. For the nuclear ring of
NGC 6951, for instance, \citet{maz08} derived the current $\SFR \sim
1.4\Aunit$ based on the old H$\alpha$--SFR relation of
\citet{ken98}, which is known to yield a larger SFR by a factor of
$\sim1.47$ than the newly calibrated relation (e.g.,
\citealt{hao11,ken12}). Also, local variations in the stellar-age
mix, initial mass function, gas/dust geometry, etc. are likely to
contaminate the derived SFRs to some extent \citep{ken12}. On the
other hand, \citet{van13} measured the ages and masses of stellar
clusters directly to obtain a temporal history of SFR in the ring of
the same galaxy. They found $\SFR \sim 0 - 0.15\Aunit$ during the
past 1 Gyr, with the current SFR less than $0.03\Aunit$, much
smaller than value reported by \citet{maz08}. Regarding the pattern
speeds, a model-independent kinematic method proposed by
\citet{tre84} has been widely used to measure the angular velocities
of arms and bars (e.g.,
\citealt{zim04,ran04,mer06,mei08,fat09,spe11}). This method relies
critically on a few assumptions, notably that a galactic disk is in
a steady state and that there is a well-defined pattern, the
validity of which is not always guaranteed.  For instance, star
formation and ensuing feedback make the density and velocity fields
non-steady in a galactic disk. When arms are not corotating with a
bar, there are significant non-steady motions in the gas flows in
the region where the bar joins the arms, which is likely to
compromise the derived pattern speeds based on gas tracers.

In addition to pattern speeds, there are many other factors such as
the ring size, bar strength, magnetic fields that may affect the
ring SFR.  Therefore, it is not yet viable to make a definitive
comparison of SFRs between our numerical predictions and
observations. Nevertheless, the observational data tabulated in
Table 1 of \citet{maz08} show that the averaged ring SFR of SAB and
SB galaxies are $2.90\pm 2.00\Aunit$ and $2.00\pm1.71\Aunit$,
respectively. Since the relative importance of spiral arms is larger
for SAB than SB galaxies, these observations are not inconsistent
with the idea of arm-enhanced SFR in the nuclear rings. In addition,
the spiral arms of NGC 4314 seem to corotate with the bar
\citep{but09}, and the ring SFR in this galaxy is quite low at
$\sim0.1\Aunit$ \citep{maz08}. On the other hand, NGC 4321 known to
undergo starburst activities in the ring \citep{ryd99} have spiral
arms rotating slower than the bar \citep{her05}. These are
consistent with our result that the arm pattern speed affects the
ring SFR.

To explore star formation in nuclear rings, we have adopted a very
simplified model of gas in barred-spiral galaxies. First of all, we
treated the gas as being isothermal and unmagnetized, whereas the
interstellar gas in real disk galaxies is multi-phase, magnetized,
and turbulent (e.g., \citealt{wol03,mck07}). This required us to
handle star formation feedback in the form of momentum injection
rather than thermal energy injection (e.g.,
\citealt{tha01,age11,kimm14}). We considered an infinitesimally-thin
disk, which precludes a potential effect of fluid motions that
involve the vertical direction.  Most importantly, we here adopted a
simple bar potential with fixed strength and pattern speed. Recent
$N$-body simulations for bar formation show that not only the bar
strength but also the bar size and pattern speed vary with time over
a few Gyrs (e.g., \citealt{min12,man14}). The parameters of spiral
arms also appear to change as a bar evolves (e.g.,
\citealt{ath12,roc13}). Therefore, it would be interesting to study
how star formation in nuclear rings studied in this work would
change in a more realistic environment where a bar, consisting of
live stellar particles, is self-generated and interacts with the
gaseous component under radiative cooling and heating, which would
be an important direction of future research.

\acknowledgments We gratefully acknowledge an insightful report form
the referee. This work was supported by the National Research
Foundation of Korea (NRF) grant, No.\ 2008-0060544, funded by the
Korea government (MSIP). A part of this work was conducted while
participating in the program entitled ``Gravity's Loyal Opposition:
The Physics of Star Formation Feedback" held at the KITP, Santa
Barbara, USA, which was supported in part by the National Science
Foundation under Grant No.\ NSF PHY11-25915.

\clearpage
\newpage

\end{document}